# Reversible Data hiding in Encrypted Domain with Public Key Embedding Mechanism

Yan Ke, Minqing Zhang, Xinpeng Zhang, *Member IEEE*, Yiliang Han, Jia Liu

*Abstract*— Considering the prospects of public key embedding (PKE) mechanism in active forensics on the integrity or identity of ciphertext for distributed deep learning security, two reversible data hiding in encrypted domain (RDH-ED) algorithms with PKE mechanism are proposed, in which all the elements of the embedding function shall be open to the public, while the extraction function could be performed only by legitimate users. The first algorithm is difference expansion in single bit encrypted domain (DE-SBED), which is optimized from the homomorphic embedding framework based on the bit operations of DE in spatial domain. DE-SBED is suitable for the ciphertext of images encrypted from any single bit encryption and learning with errors (LWE) encryption is selected in this paper. Pixel value ordering is introduced to reduce the distortion of decryption and improve the embedding rates (ER). To apply to more flexible applications, public key recoding on encryption redundancy (PKR-ER) algorithm is proposed. Public embedding key is constructed by recoding on the redundancy from the probabilistic decryption of LWE. It is suitable for any plaintext regardless of the type of medium or the content. By setting different quantization rules for recoding, decryption and extraction functions are separable. No distortion exists in the directly decrypted results of the marked ciphertext and ER could reach over 1.0 bits per bit of plaintext. Correctness and security of the algorithms are proved theoretically by deducing the probability distributions of ciphertext and quantization variable. Experimental results demonstrate the performances in correctness, one-way attribute of security and efficiency of the algorithms.

*Index Terms*—Information hiding, reversible data hiding in encrypted domain, learning with errors, difference expansion.

## I. INTRODUCTION

REVERSIBLE data hiding in encrypted domain (RDH-ED) is an information hiding technique that aims to not only accurately embed and extract the additional data in the ciphertext, but also restore the original plaintext losslessly [1][2]. RDH-ED is useful in some distortion intolerable applications, such as ciphertext management or retrieval in the cloud or telemedicine, and active forensics on the integrity or identity of ciphertext in secret communication systems. With the increasing demand for information security and the development of the encrypted signal processing techniques, RDH-ED has been an issue of great attention in the fields of privacy protection and encrypted signal processing [2].

The application scenarios of RDH-ED technology are mainly derived from the applications of cryptosystems in practice.

This work was supported in part by National Natural Science Foundation of China under Grant 61872384 and Grant 62102450.

Yan Ke is with Engineering University of PAP and Research Institute of Hi-Tech, Xi'an, 710086, China (e-mail: 15114873390@163.com).

Minqing Zhang, Yiliang Han, and Jia Liu are with the School of Cryptography Engineering in Engineering University of PAP, Xi'an, 710086, China (e-mail: api_zmq@126.com; hanyil@163.com; liujia1022@gmail.com).

Xinpeng Zhang is with School of Computer Science, Fudan University, Shanghai, 200000, China (e-mail: zhangxinpeng@fudan.com).

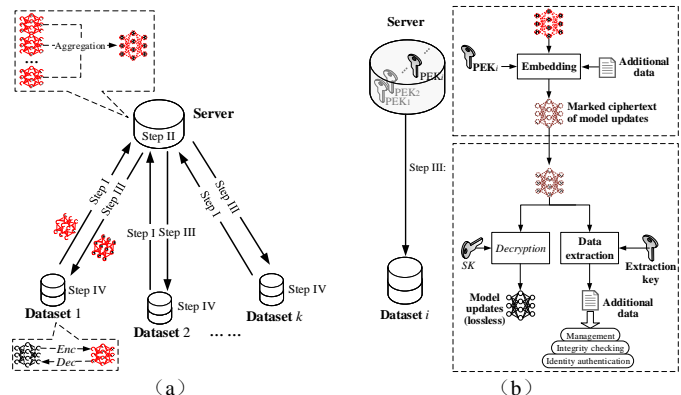

Fig. 1. The application instance of PKRDH-ED in federated learning: (a) Architecture of horizontal federated learning system; (b) Application.

Recently, researches of RDH-ED have made extraordinary advancements, in terms of embedding capacity (EC), security, separability, etc. However, the application scenarios of most algorithms are still too insufficient to meet the needs of the developing environment of ciphertext. Novel applications of cryptosystems have been emerging, such as federated learning, secure multi-party computing (MPC), and block chain. The requirements of authentication, and management in the applications of key generation, ciphertext transmission, etc. have showed a trend of diversification, dynamic, and personalization. This provides new directions and technical challenges for the research of RDH-ED [3].

Existing researches of RDH-ED concentrate on constructing algorithms under *the symmetric system*, such as those based on stream encryption [1][4]-[8], advanced encryption standard (AES) [9][10], and RC4 encryption [11]. The characteristic of the symmetric algorithm is that the data hider and data extractor share the same key or no key is needed. There are limitations in the application of secret communication scenarios based on public key cryptosystems or multi-party secure computing systems due to the contradiction in key distribution. Therefore, it is urgent to construct the public key mechanism of reversible data hiding in encrypted domain. In this paper, we first discuss the characteristics, application scenarios, and security requirements of *public key embedding* (PKE) mechanism of RDH-ED technology.

Drawing on the public key cryptosystems, the main characteristics of public key reversible data hiding in encrypted domain (PKRDH-ED) is that the key for embedding should be different from the key for extraction. And it is required that all the elements of the embedding process, such as embedding algorithm, parameter setting, application interface or embedding key, can be publicly available to all (trusted or untrusted) nodes in the communication network, while the extraction process should be kept secret to public and be performed only by legitimate private users.

Constructing PKE mechanism of RDH-ED is of great significance to improve the practicability of RDH-ED technology in the application of ciphertext authentication and management in public key cryptosystems, MPC and distributed deep learning privacy.

In Fig. 1, we illustrate the application of PKRDH-ED in horizontal federated learning system. Federated learning was proposed to build machine learning models based on distributed datasets with the help of a (untrusted) cloud server [12][13]. To prevent privacy leakage to the server, $k$ participants with the same data structure collaboratively learn a machine learning model via homomorphic encryption [14]. There are four steps in horizontal federated learning as shown in Fig. 1(a). *Step I*: participants locally compute training gradients, encrypt them and send the ciphertext to server. *Step II*: server performs secure aggregation without learning any information about the gradients and participants. *Step III*: server send back the aggregated ciphertext of gradients to participants for update. *Step IV*: participants decrypt the gradients and update their respective model.

In the above system, a typical assumption is that the participants are honest whereas the server is untrusted and no leakage of information from any participants to the server is allowed. However, due to the lack of effective authentication on the origin and integrity of ciphertext in *Step III*, an attacker (or malicious participant) can pretend to be a server to return customized ciphertext of gradients to participant $i$. Then poisoning attacks [15], adversarial examples attack [16] or backdoor attacks [17] might be implemented on the model update of participant $i$ based on dataset $i$. Faced with the privacy risks, as Fig.1(b) shows, the legitimate server can use PKRDH-ED in *Step III* to embed additional data into the ciphertext for management, identity authentication or integrity verification, thus improving the security of model update in further steps.

Besides, to improve the model update efficiency in *Step IV*, personalized feedbacks derived from the effect of secure aggregation in *Step II* or feedbacks from users in application shall be returned to participant $i$ in *Step III* [13]. PKRDH-ED provides a secure method of returning feedbacks together with ciphertext, thus improve the efficiency of model update and secure aggregation in further steps.

In the application of PKRDH-ED, a sender embeds the information about ciphertext management or authentication into the ciphertext by using the *public embedding key* (PEK) and transmits the marked ciphertext. The receiver owns the extraction key in secret and he can extract information by using the extraction key. The advantages of such a method include: *a)* The additional data used for authentication or management and the to-be-authenticated ciphertext are fused into one by the method of embedding, which realizes the direct binding of the two types of data and can resist attacks such as tailors, forgery, and tampering. *b)* The traffic of confidential data transmission are lessened in the authentication communication. *c)* The PEK mechanism of RDH-ED can effectively reduce the total amount of keys in an RDH-ED system from $O(n^2)$ to $O(n)$, thus improving the efficiency of key distribution and update.

The instance in Fig. 1 can also be applied to vertical federated learning system. The reversibility of RDH-ED ensures the lossless decryption and availability of the ciphertext. The functions of PKRDH-ED in the security and efficiency improvements of federated learning relies on PKE mechanism, that is, the embedding process can be implemented publicly, while other individuals other than the legitimate extractor (participant $i$) cannot gain any information from the marked ciphertext. Therefore, different from the existing symmetric RDH-ED technology, the security of PKRDH-ED not only requires the consistent security of embedding and the confidentiality of the embedded information [3][18], but also requires the *one-way attribute* of public information, *e.g.*, PEK. PEK should not reveal any information of deducing the extraction key.

In summary, PKRDH-ED can be used to design complex, dynamic and personalized security authentication interaction protocols, and can provide technical support for ciphertext authentication and management in multi-party or multi-authority scenarios.

## II. RELATED WORK

The cryptosystems introduced into RDH-ED mainly include symmetric encryption, public key encryption, and secret sharing. The characteristic of public key encryption is that anyone could get access to the public key and encrypt information, and the only the secret key owner can decrypt the information. The relationship between the secret communication parties is asymmetric. Similar to public key encryption, public key RDH-ED actually aims to construct an asymmetric secret communication based on the operation of information embedding and extraction. Therefore, various practical applications of public key encryption (such as the federated learning based on homomorphic encryption) can provide wide application prospects for PKRDH-ED technology. The construction of PKRDH-ED should comprehensively consider the application of public key cryptography and the requirements on RDH-ED. Related works of RDH-ED are as follows:

The methodologies of RDH-ED algorithm can be classified into three categories: "vacating room after encryption (VRAE)"[4]-[7], "vacating room before encryption (VRBE)" [1][8]-[10] (room, namely the redundancy in the cover, is vacated for data hiding in encrypted domain), and "vacating redundancy in encryption (VRIE)[19][20]". Puteaux introduced AES into RDH-ED [9]. Zhang first proposed a stream cipher based VRAE method for encrypted images [4]. Then separable VRAE RDH-ED was proposed in [22][23]. Puteaux [24] proposed to make full use of prediction error before encryption in a VRBE algorithm with high embedding capacity (EC). To preserve as much correlation as possible after encryption, a novel VRAE framework with a specific stream cipher was proposed via reusing the same key within a pixel block [25]. It could provide certain security guarantees, but key reusing would weaken the encryption intensity of symmetric encryption in theory. Since it is difficult to vacate room after encryption, current attentions focus more on VEBE methods [8][24], while more computational expense is introduced into the client end.

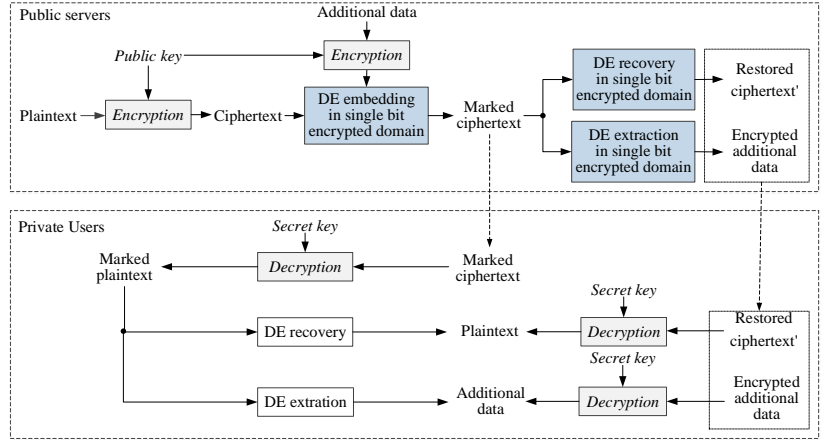

Fig. 2. Framework of DE-SBED.

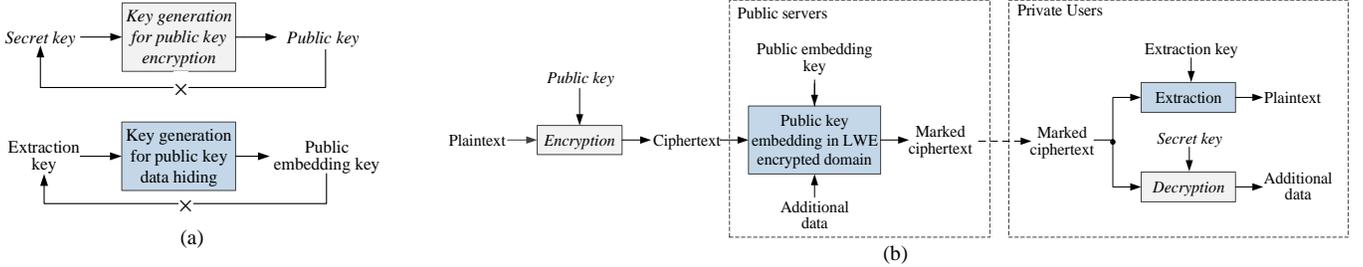

Fig. 3. Framework of PKR-ER: (a) Key generation; (b) Framework.

The above algorithms are mainly based on symmetric encryption. Researches of public key encryption based RDH-ED are mainly based on Paillier encryption [26]-[32] and learning with errors (LWE) encryption [19]-[21]. Probabilistic and homomorphic properties of the above cryptography allow the third party, *i.e.*, the cloud servers, to conduct operations directly on ciphertext without knowing the secret key, which shows potential for more flexible realizations of RDH-ED. The first Paillier encryption based RDH-ED was proposed by Chen *et al.* [26]. Shiu *et al.* [27] and Wu *et al.* [28] improved the EC of [26] by solving the pixel overflow problem. Those algorithms were VRBE methods. Li *et al.* in [31] proposed a VRAE method with a considerable EC by utilizing the homomorphic addition property of Paillier encryption and HS technique. Wu *et al.* proposed two RDH-ED algorithms for the encrypted images in [30]. Zhang *et al.* [29] proposed a combined scheme consisting of a lossless scheme and a reversible scheme to realize separable RDH-ED. Xiang vacated room before encryption by using RDH [34] and then embedded the additional data into the ciphertext of vacated LSBs by employing homomorphic multiplication.

Different from VRAE or VRBE, VRIE aims to utilize the redundancy generated by the encryption process for embedding [20]. The ciphertext extension produced by public key encryption can provide a large amount of ciphertext redundancy for embedding. More efficient methods can be achieved by re-quantizating the redundant domain and recoding the ciphertext [19][20]. LWE based RDH-ED was first proposed in [19] by quantifying the LWE encrypted domain and recoding the redundant ciphertext. Ke *et al.* fixed the parameters for LWE encryption and proposed a multi-level RDH-ED with a flexible applicability and high EC in [20]. Fully homomorphic encryption (FHE) is based on the redundancy of ciphertext data structure, and can be used for constructing flexible and complex operations of embedding. Ke *et al.* in [3] introduced FHE to encapsulate the RDH method of difference expansion (DE) [35]. It has advantages in ensuring security, reversibility and separability, but the computational complexity is high.

In summary, the public key cryptosystems and their applications are suitable for PKRDH-ED. VRIE is feasible for constructing public key embedding because it could take advantage of the redundancy from ciphertext expansion. Specifically, in this paper, we choose lattice based LWE public key encryption to construct PKRDH-ED algorithms. There are four advantages of LWE encryption: *a)* It has reliable theoretical security that could resist anti-quantum algorithm analysis. Lattice based encryption is the main candidate of post-quantum cryptosystems in the future. *b)* It has high running speed due to its linear structure and operations in lattice. *c)* The ciphertext extension of lattice ciphers can provide sufficient ciphertext redundancy for embedding. *d)* Lattice based cipher is currently the only cryptosystem that supports to construct FHE, zero-knowledge proof and secret sharing cryptosystems, which plays an important role in federated learning and secure MPC. Therefore, it provides considerable potential applications of ciphertext authentication and management for PKRDH-ED.

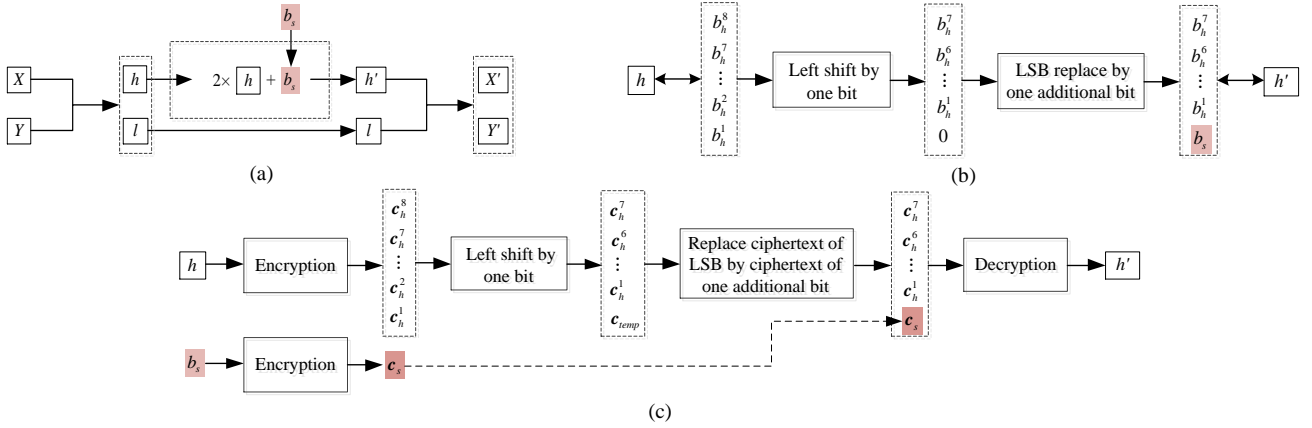
Fig. 4. The analogy and contrast of DE in different domains: (a) In spatial domain; (b) In bit-spatial domain; (c) In LWE encrypted domain.

The rest of this paper is organized as follows. The following section introduces the frameworks and the detailed processes of the proposed two algorithms. In Section IV, the three judging standards of RDH-ED, including *correctness*, *security* and *efficiency*, are discussed theoretically and verified with experimental results. Finally, Section V summarizes the paper and discusses future investigations.

### III. THE PROPOSED SCHEMES

In this paper, we focus on the realization of PKE mechanism. Two PKRDH-ED algorithms with different PKE applications are proposed: *difference expansion in single bit encrypted domain (DE-SBED)* and *public key recoding on encryption redundancy (PKR-ER)*.

#### A. Frameworks

**DE-SBED**: PKE mechanism is realized based on the framework of fully homomorphic encryption encapsulated difference expansion (FHEE-DE) in [3]. Actually, the homomorphic embedding operations based on FHE could satisfy the requirements of the public key mechanism, because it can support any third party in public to implement homomorphic embedding while ensuring the privacy and security of the plaintext. Only the secret key owner can decrypt the ciphertext and extract data from the marked plaintext. FHEE-ED has provided a high secure PKRDH-ED instance, but the computational complexity of FHE is too high to be practical. Following the framework of FHEE-DE, DE-SBED is proposed based on the characteristics of single-bit encryption of LWE. The computational complexity has been effectively reduced. Fidelity and embedding capacity have also been improved by introducing pixel value ordering (PVO) technology. It should be noted that DE-SBED is suitable for any single bit encryption algorithm, and we select LWE encryption in this paper.

The framework of DE-SBED is as shown in Fig. 2. The plaintext is encrypted into ciphertext via *public key*. The public server first encrypts the additional data via *public key* and then performs *DE embedding in single bit encrypted domain* to obtain the marked ciphertext. Then *a)* the user could directly decrypt the marked ciphertext via *secret key* to obtain the marked plaintext. *DE extraction or recovery* can be implemented on the marked plaintext to obtain the additional data or plaintext; *b)* the server could restore the ciphertext by performing *DE recovery in single bit encrypted domain* on the marked ciphertext. The restored ciphertext can be losslessly decrypted by the user via *secret key*; *c)* the server could also return the encrypted additional data by performing *DE extraction in single bit encrypted domain* on the marked ciphertext. Then the user could decrypt the encrypted additional data via *secret key*.

**PKR-ER**: To satisfy the needs of more flexible applications of PKRDH-ED, public embedding key (PEK) is constructed based on the redundancy in encryption of LWE. Therefore, it has nothing to do with the type of medium or the content of the plaintext. PEK is independent from the public key for encryption. Any third party could have access to PEK while only the secret key owner could extract information from the marked ciphertext or decrypt it.

The framework of PER-ER is as shown in Fig. 3. PKE mechanism requires the public embedding key generation function to be one-way as shown in Fig. 3(a). The plaintext is encrypted into ciphertext via *public key*. Additional data is embedded into ciphertext to obtain the marked ciphertext by the public server via *public embedding key*. With the marked ciphertext, the user could extract the additional data from it; or he could directly decrypt it to obtain the plaintext via *secret key*.

#### B. Difference Expansion in Single Bit Encrypted Domain

*1) Methodology*

The notation of the main variables is shown in Table I.

We illustrate the methodology of DE-SBED by the analogy and contrast of DE in different domains as shown in Fig. 4. Tian's DE algorithm [35] is as shown in Fig. 4(a): two adjacent pixels $X$ and $Y$ from an image $I$ can be used to hide one bit data $b_s$, where $0 \leqslant X, Y \leqslant 255$ and $b_s \in \{0, 1\}$. The difference $h$ and average value $l$ (integer) of $X$ and $Y$ are computed as following:

$$h = X - Y \qquad (1)$$

$$l = \left\lfloor \frac{X+Y}{2} \right\rfloor \qquad (2)$$

$$X = l + \left\lfloor \frac{h+1}{2} \right\rfloor \qquad (3)$$

$$Y = l - \lfloor h/2 \rfloor \qquad (4)$$

TABLE I
KEY DISTRIBUTION

| Denotation | Representation |
|---|---|
| $X, Y \in Z_{256}$ | A pair of two adjacent pixels |
| $h$ | The difference between $X$ and $Y$. |
| $l$ | The mean of $X$ and $Y$. |
| $b_h^i, b_l^i \in \{0,1\}$ | The $i$-LSB of $h$ or $l$ ($i$=1, 2, …, 8). |
| $c_h^i, c_l^i \in Z_q^n$ | The ciphertext of $b_h^i$ or $b_l^i$ ($i$=1, 2, …, 8). |
| $b_s \in \{0,1\}$ | The additional bit to be embedded. |
| $c_{bs}$ | The ciphertext encrypted from $b_s$. |

Assuming $X > Y$, and $\lfloor . \rfloor$ is the floor function meaning "the biggest integer less than or equal to" while $\lceil . \rceil$ is the ceiling function.

Embedding:
$$h' = 2 \times h + b_s \quad (5)$$
The embedded pixels $X'$ and $Y'$ can be obtained by substituting $h'$ into Eqs. (3), (4).

Extration:
$$b_s = LSB(h') \quad (6)$$
$LSB(.)$ is to obtain the least significant bit of the input integer.

Recovery:
$$h = \lfloor h'/2 \rfloor \quad (7)$$
Then, the original pixels $X$ and $Y$ can be recovered by using Eqs. (3), (4).

In Tian' algorithm, the embedding is mainly implemented on $h$, the pixel difference value. Then we convert all the embedding operation in Eq. (5) into bit operations as shown in Figure 4(b). The bit operations are mainly *left shift* and *bit replacement*. Since the LWE algorithm is the single bit encryption, the difference expansion can be achieved in the encrypted domain by operating ciphertext position left shifting and replacement as shown in Fig. 4(c). To improve EC and reduce the distortion of directly decrypted result from the marked ciphertext, we sort the image pixels before encryption.

2) *Preprocessing*
   a) *Pixels value ordering*

The plaintext is a 512×512 image $I$. For each row, the pixels are scanned as $(p_1, p_2, …, p_l)$, $l$=512. Then the sequence is sorted in a descending order as $\{p_{\sigma(1)}, p_{\sigma(2)}, …, p_{\sigma(l)}\}$, where $\sigma$: $\{1,2,…,l\} \rightarrow \{1,2,…,l\}$ denotes the unique one-to-one mapping such that $p_{\sigma(i)} \leqslant p_{\sigma(j)}$, if $i > j$. The sorted plaintext $I'$ is obtained.

   b) *Constraints*

$I'$ is divided into non-overlapping pixel pairs. Each pair consists of two adjacent pixels $(X, Y)$, where $0 \leqslant X, Y \leqslant 255$. As grayscale values are bounded in [0, 255], we have constraints about $h$ and $l$ according to Eqs. (1), (2):
$$0 \leqslant l + \lfloor \frac{h+1}{2} \rfloor \leqslant 255 \quad (8)$$
$$0 \leqslant l - \lfloor \frac{h}{2} \rfloor \leqslant 255 \quad (9)$$

To avoid overflow or underflow problems, we use a map matrix $M_{ava} \in \{0,1\}^{256 \times 256}$ to indicate available pixel pairs. Value "1" indicates the bigger pixel within an available adjacent pixel pair for DE data hiding. The difference $h$ of an available pair should satisfy the following constraints [35]:
$$|h| \leqslant \min(2(255-l), 2l+1) \quad (10)$$
$$|2 \cdot h + b_s| \leqslant \min(2(255-l), 2l+1) \quad (11)$$
for $b$= 0 or 1.

We add an extra fidelity constraint: the available pixel pairs are preferentially selected with a smaller pixel difference. The fidelity parameter $h_{fid}$ is introduced:
$$h \leqslant h_{fid} \quad (12)$$
$M_{ava}$ would be lossless compressed as side information of the ciphertext to superimpose on the host signal. In this paper, we have used key-switching based LSB (KS-LSB) [3] to embed $\sigma$ and $M_{ava}$ into the ciphertext in the experiments.

   c) *Parameters setting and function definition*

The cryptosystem is parameterized by the following [36]: $n$ (the length of the secret key), $q$ (the minimum prime between $n^2$ and $2n^2$), $\mathbb{Z}_q$ (the modulus of a finite field), $d \geqslant (1+\varepsilon)(1+n)\log_2 q$ (the dimension of the public key space), $\varepsilon > 0$. If $q$ is a prime, all the operations are performed modulo $q$ in $\mathbb{Z}_q$. We denote the noise probability distribution on $\mathbb{Z}_q$ as $\chi$, $\chi = \overline{\Psi}_{\alpha q}$ which is a discrete Gaussian distribution:
$$\overline{\Psi}_{\alpha q} = \{\lceil qx \rfloor \bmod q \mid x \sim N(0, \alpha^2)\} \quad (13)$$
and $\lceil qx \rfloor$ denotes rounding $qx$ to the nearest integer [20].

*Definition 1* [37]: The secret key of LWE encryption generation:
$$s = SKGen_{(n,q)}(.) \quad (14)$$
in which the secret key $s \in \mathbb{Z}_q^n$ is sampled independently and uniformly.

*Definition 2* [37]: The public key of LWE encryption generation:
$$(A, P) = PKGen_{(d, n, q)}(s) \quad (15)$$
in which a matrix $A \in \mathbb{Z}_q^{n \times d}$ is generated uniformly and a $d$-dimension vector $e \in \mathbb{Z}_q^d$ is sampled from the distribution $\chi$, then the vector $p \in \mathbb{Z}_q^d$ is obtained:
$$p = A^T \cdot s + e \quad (16)$$

*Definition 3* [37]: The encrypting function:
$$c = Enc_{(A,P)}(m) \quad (17)$$
which returns a vector $c = (u, c) \in \mathbb{Z}_q^n \times \mathbb{Z}_q$ as the ciphertext of $m \in \mathbb{Z}_2$. Generate a random vector $a_r \in Z_2^d$ uniformly and output $u \in \mathbb{Z}_q^n$ and $c \in \mathbb{Z}_q$:
$$u = A \cdot a_r \quad (18)$$
$$c = p^T \cdot a_r + m \cdot \lfloor q/2 \rfloor \quad (19)$$

*Definition 4* [37]: The decrypting function:
$$m = Dec_{(s)}(c) \quad (20)$$
which returns the decrypted bit $m \in \{0, 1\}$ via $s$. Calculated a quantization variable $\lambda \in \mathbb{Z}_q$ and then $m$ could be obtained:
$$\lambda = c - s^T \cdot u \quad (21)$$

TABLE II
KEY DISTRIBUTION IN PKRDH-SBED

| Classification | Denotation | Function | Owner |
|---|---|---|---|
| Public key | ($A$, $P$) | Encryption. | Open to the public |
| Secret key | $s$ | Decryption. | Private user |

$$m = \begin{cases} 0, \lambda \in [0, \lfloor q/4 \rfloor) \cup [\lfloor 3q/4 \rfloor, q) \\ 1, \lambda \in [\lfloor q/4 \rfloor, \lfloor 3q/4 \rfloor) \end{cases} \quad (22)$$

*3) Key Distribution*

Different keys are distributed as shown in Table II:

*4) Encryption*

For the pixel pair ($X$, $Y$), ($h$, $l$) of ($X$, $Y$) are first calculated Then, ($h$, $l$) instead of ($X$, $Y$) would be encrypted as ciphertext: $c_h^i = Enc(b_h^i)$ and $c_l^i = Enc(b_l^i)$ ($i$=1, 2, … , 8). We omit the symbol "$_{(A, P)}$" in Eq. (17) for short in the following.

*5) DE Embedding in Single Bit Encrypted Domain*

Step 1: The ciphertext are ($c_h^8, c_h^7, …, c_h^1$) and ($c_l^8, c_l^7, …, c_l^1$). Calculate $c_{temp0} = Enc(0)$. One position left shift is implemented on ($c_h^8, c_h^7, …, c_h^1$) to obtain the ciphertext of expanded $h$: ($c_h^7, c_h^6, …, c_h^1, c_{temp0}$).

Step 2: Calculate $c_{bs} = Enc(b_s)$. Replace ciphertext of the LSB by the ciphertext of the additional bit to obtain the ciphertext of DE embedded $h'$: ($c_{h'}^8, c_{h'}^7, …, c_{h'}^1$) = ($c_h^7, c_h^6, …, c_h^1, c_{bs}$).

*6) Data Extraction and Plaintext Recovery by Private Users*

The user owns the secret key $s$ for decryption. After receiving the marked ciphertext, the user could decrypt the marked ciphertext to obtain $h'$ and $l$ via $s$: $b_{h'}^i = Dec_{(s)}(c_{h'}^i)$, $b_l^i = Dec_{(s)}(c_l^i)$, ($i$=1, 2, … , 8).

The additional data could be extracted according to DE extraction in Eq. (6). $h$ and ($X$, $Y$) could be recovered according to DE recovery in Eqs. (7), (3)-(4).

*7) DE Recovery & DE Extraction in Single Bit Encrypted Domain by Public Servers*

In public environments, no information about decryption could be gathered. With the marked ciphertext, the servers could restore a new ciphertext of the plaintext.

Step 1: Calculate $c_{temp0} = Enc(0)$.

Step 2: One position right shift is implemented on ($c_{h'}^8, c_{h'}^7, …, c_{h'}^1$) to obtain the unexpanded ciphertext of $h$: ($c_{temp0}, c_{h'}^8, c_{h'}^7, …, c_{h'}^2$).

The restored ciphertext is obtained: ($c_{temp0}, c_{h'}^8, c_{h'}^7, …, c_{h'}^2$) and ($c_l^8, c_l^7, …, c_l^1$).

After receiving the restored ciphertext, the user could implement the decryption to obtain the original $h$ and $l$ via $s$: $b_h^i = Dec_{(s)}(c_h^i)$, $b_l^i = Dec_{(s)}(c_l^i)$, ($i$=1, 2, …, 8).

The encrypted $b_s$ is $c_{h'}^1$, which can be extracted from ($c_{h'}^8, c_{h'}^7, …, c_{h'}^1$). The encrypted additional data could also be decrypted by the user via the secret key $s$: $b_s = Dec_{(s)}(c_{h'}^1)$.

*C. Public Key Recoding on Encryption Redundancy*

*1) Methodology*

PKE mechanism is constructed by recoding on the redundancy from ciphertext extension, which comes from the fault tolerance of the mapping relationship in the probabilistic decryption process of LWE in Eq. (21). The process of LWE decryption is analyzed as following:

$$\lambda = c - s^T \cdot u = p^T \cdot a_r + m \cdot \lfloor q/2 \rfloor - s^T \cdot A \cdot a_r = (A^T \cdot s + e)^T \cdot a_r - s^T \cdot A \cdot a_r + m \cdot \lfloor q/2 \rfloor = e^T \cdot a_r + m \cdot \lfloor q/2 \rfloor \quad (23)$$

Without additive noise, i.e., $e^T \cdot a_r = 0$, $\lambda$ in Eq. (23) would be 0 if $m$=0, or $\lfloor q/2 \rfloor$ if $m$=1. Introducing additive noise, i.e., $e^T \cdot a_r \neq 0$, the value of $h$ is fluctuant around 0 if $m$=0 or $\lfloor q/2 \rfloor$ if $m$=1. By controlling $\alpha^2$, the variance of the distribution $\chi$, the fluctuation interval of $\lambda$ can be controlled not to exceed $\lfloor q/4 \rfloor$. As Fig. 5(a) shows, we represent integers in $\mathbb{Z}_q$ with the points of a circle. $\mathbb{Z}_q$ is equally divided into 4 regions (I, II, III, and IV). According to Eq. (22), when $\lambda$ is located in the regions I and IV, namely $\lambda \in [0, \lfloor q/4 \rfloor) \cup [\lfloor 3q/4 \rfloor, q)$, the decrypted result is 0; when $\lambda$ is located in the regions I and IV, namely $\lambda \in [\lfloor q/4 \rfloor, \lfloor 3q/4 \rfloor)$, the decrypted result is 1.

Without additive noise, finding the secret key $s$ with the public key ($A$, $P$) would be easy: via the $d$ equations in $P$, we can recover $s$ in polynomial time by using the Gaussian elimination algorithm. Introducing additive noise, Gaussian elimination algorithm would amplify the noise to an unmanageable level and leave essentially no information about $s$ in the elimination results. The best-known cryptanalysis algorithms for LWE run in exponential time (even quantum algorithms do not appear to help) [38].

In fact, we would like to take advantage of the redundant decryption resulted by the additive noise to embed additional data. Fig. 5(b) shows the methodology of embedding one bit of additional data:

Each region is equally divided into 2 sub-regions, denoted as I.0, I.1; II.0, II.1; III.0, III.1; and IV.0, IV.1. By controlling $\alpha^2$, the fluctuation range of $\lambda$ is limited to no more than $\lfloor q/8 \rfloor$, i.e., $|e^T \cdot a_r| < \lfloor q/8 \rfloor$ (In practice, $\lambda$ is even limited to no more than $\lfloor q/10^2 \rfloor$ to ensure the correct decryption). Then $h$ would be located in the "0" sub-region, i.e., I.0 or IV.0 (if $m$=0); or in II.0 or III.0 (if $m$=1). When the additional bit is "0", the ciphertext remains unchanged; when the additional bit is "1", the position of $\lambda$ is changed to the "1" sub-region within the same region by adding or subtracting a quantization step, $\lfloor q/8 \rfloor$, to the ciphertext. An instance is as shown in Fig. 5(b): the original $h$ is point B located in II.0. To embedding the bit "1", the marked $h'$ is B' located in II.1.

Next, we detail PKR-ER. Since the embedding is implemented based on redundancy in the encryption, it has nothing to do with the content of the plaintext. Therefore, we detail the processes of embedding additional data into ciphertext of one bit of plaintext. The plaintext is denoted as $m$

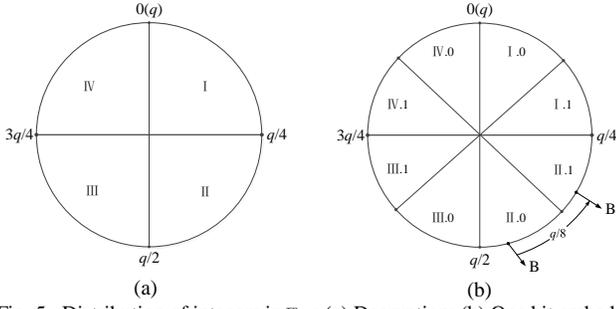

Fig. 5. Distribution of integers in $\mathbb{Z}_q$: (a) Decryption; (b) One bit embedding.

TABLE III
KEY DISTRIBUTION IN PKRDH-ED ALGORITM 2

| Classification | Denotation | Function | Owner |
|---|---|---|---|
| Public key | $(A, P)$ | Encryption. | Open to the public |
| Public embedding key | $\gamma Q_{step}$ | Embedding | Open to the public |
| Secret key | $s$ | Decryption and Data extraction | Private user |

$\in \mathbb{Z}_2$. The to-be-embedded data is an *N*-bit message, denoted as $\boldsymbol{b}_e = \{b_e(1), b_e(2),\ldots, b_e(N)\} \in \mathbb{Z}_2^N$. The quantization step is denoted as $Q_{step}$, $Q_{step} = \lfloor q/2^{N+2} \rfloor$. Each region would be equally divided into $2^N$ sub-regions, denoted as I.0, I.1, I.2,…, I.$2^N$-1; II.0,…, II.$2^N$-1; III.0,…, III.$2^N$-1; IV.0, …, IV. $2^N$-1.

*2) Secret key/Public Key Generation and Encryption*

All the operations in this step are the same with the operations in *PKRDH-ED based on single bit encrypted domain*. It should be noted that to ensure the correct decryption of the marked ciphertext, the fluctuation range of $\lambda$ is limited to no more than $Q_{step}$, i.e., $|\boldsymbol{e}^T \cdot \boldsymbol{a}_r| < Q_{step}$.

Secret key: $\boldsymbol{s} = SKGen_{(n, q)}(.)$.
Public key: $(\boldsymbol{A}, \boldsymbol{P}) = PKGen_{(d, n, q)}(\boldsymbol{s})$.
Encryption: $\boldsymbol{c} = Enc(m)$.

*3) Public Embedding Key and Key Distribution*

Calculate the quantization variable $\lambda$ according to Eq. (21). Then the sign factor $\gamma$ is obtained:

$$\gamma = \begin{cases} +1, \lambda \in [0, \lfloor q/4 \rfloor) \cup [\lfloor q/2 \rfloor, \lfloor 3q/4 \rfloor) \\ -1, \lambda \in [\lfloor q/4 \rfloor, \lfloor q/2 \rfloor) \cup [\lfloor 3q/4 \rfloor, q) \end{cases} \quad (24)$$

The sign factor determines the direction of ciphertext' changing when embedding. Public embedding key includes two parts: the sign factor and the quantization step. Keys are distributed in Table III.

*4) Embedding*

The (untrusted) server could implement additional data embedding into the ciphertext via *public embedding key* in public environment.

Step 1: Convert the *N*-bit message $\boldsymbol{b}_e$ into decimal data as the to-be-embedded data $m_e \in \mathbb{Z}_{2^N}$:

$$m_e = \sum_{i=1}^{N} b_e(i) \times 2^{i-1} \quad (25)$$

Step 2: Embed $m_e$ into $c$ by recoding the ciphertext to obtain the marked ciphertext $c'$:

$$c' = c + m_e \times \gamma Q_{step} \quad (26)$$

The marked ciphertext $\boldsymbol{c}'$ is $(\boldsymbol{u}, c')$.

*5) Decryption*

After receiving the marked ciphertext from the server, the private user could decrypt it to obtain the plaintext losslessly via $\boldsymbol{s}$:

$$m = Dec_s(\boldsymbol{c}') \quad (27)$$

*6) Data Extraction*

After receiving the marked ciphertext from the server, the user could extract additional data via $\boldsymbol{s}$ in private.

Step 1: the private user calculates the quantization variable $\lambda$ according to Eq. (21) via $\boldsymbol{s}$.

Step 2: Calculate the embedded data $m_e$ according to Eq. (28):

$$m_e = \begin{cases} \left\lfloor \dfrac{\lambda}{Q_{step}} \right\rfloor, \lambda \in [0, q/4) \\ 2^{N+1} - \left\lceil \dfrac{\lambda}{Q_{step}} \right\rceil, \lambda \in [q/4, q/2) \\ \left\lfloor \dfrac{\lambda}{Q_{step}} \right\rfloor - 2^{N+1}, \lambda \in [q/2, 3q/4) \\ 2^{N+2} - \left\lceil \dfrac{\lambda}{Q_{step}} \right\rceil, \lambda \in [3q/4, q) \end{cases} \quad (28)$$

Step 3: Convert $m_e$ into the binary data to obtain the additional message $\boldsymbol{b}_e$.

*D. Comparison of Applications*

There are four differences in the applications of the two algorithms:

*a)* DE-SBED is mainly for images where PVO is introduced to enhance the correlation of adjacent pixels while PKR-ER is applicable to any types of plaintext.

*b)* In DE-SBED, the server has to encrypt the additional data via the public key before embedding. Therefore, PEK is not constructed specially. In PKR-ER, PEK is constructed.

*c)* In DE-SBED, anyone could implement encryption. In PKR-ER, only the extraction owner who generates PEK could implement encryption.

*d)* The directly decrypted result of the marked ciphertext in DE-SBED contains embedded information, that is, the marked plaintext is distorted and a recovery process is needed then. In PKR-ER, the processes of decryption and extraction are independent. The direct decrypted result of the marked ciphertext is the lossless plaintext.

IV. THEORETICAL ANALYSIS AND EXPERIMENTAL RESULTS

*A. Correctness*

The correctness of PKRDH-ED schemes includes the lossless restoration of plaintext and the accurate extraction of the embedded data.

The experiments were all implemented on MATLAB2015a with a 64-bit single core i7- 10875H CPU @ 2.30GHz, 64.0 GB RAM. We use 1000 different 512×512 8-bit grayscale images from USC-SIPI (*http://sipi.usc.edu/database/database.php?volume=misc*) for testing. Six images as shown in Fig. 6 were selected to demonstrate the experimental results.

*Parameters setting*: Solving the LWE problem with given parameters is equivalent to solving Shortest Vector Problem in a lattice with a dimension $\sqrt{n\log_2(q)/\log_2(\delta)}$. Considering the efficiencies of the best known lattice reduction algorithms, the secure dimension of the lattice must reach 500 (*e.g.*, $\delta$=1.01) [39], [40]. An increase in $n$ will result in a high encryption blowup. To balance security and the efficiency, we set $n$ =240, $q$=57601, $d$ =4573. To ensure the fidelity of the marked plaintext [3], we set $h_{fid}$=10.

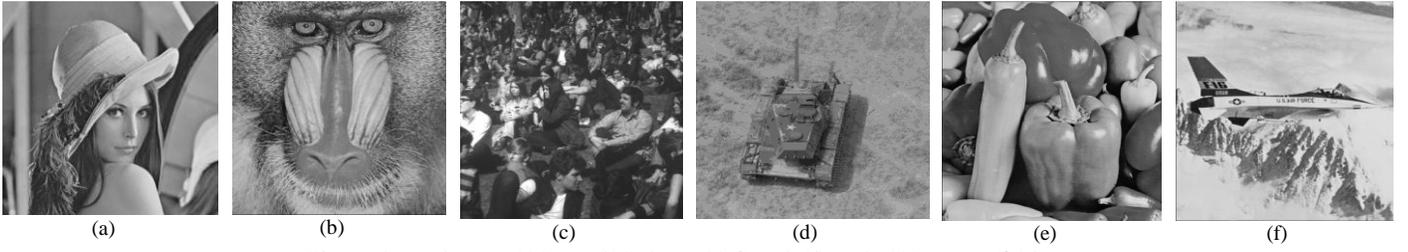

(a)     (b)     (c)     (d)     (e)     (f)

Fig. 6. The test images. (a) Lena; (b) Baboon; (c) Crowd; (d) Tank; (e) Peppers; (f) Plane.

TABLE IV
$PSNR_1$ (*DB*) VERSUS EC(*BITS*) / ER (BPP) AT DIFFERENT $h_{fid}$ OF PKRDH-ED ALGORITHM 1.

| Images | $h_{fid} = \infty$ | | $h_{fid} = 10$ | | $h_{fid} = 5$ | | $h_{fid} = 3$ | | $h_{fid} = 2$ | | $h_{fid} = 1$ | | $h_{fid} = 0$ | |
|---|---|---|---|---|---|---|---|---|---|---|---|---|---|---|
| | EC/ER | $PSNR_1$/SSIM | EC/ER | $PSNR_1$/SSIM | EC/ER | $PSNR_1$/SSIM | EC/ER | $PSNR_1$/SSIM | EC/ER | $PSNR_1$/SSIM | EC/ER | $PSNR_1$/SSIM | EC/ER | $PSNR_1$/SSIM |
| Lena | 131072 /0.5000 | 50.9706 /0.9427 | 130978 /0.4996 | 51.3121 /0.9970 | 130790 /0.4989 | 51.5099 /0.9971 | 130342 /0.4972 | 51.7181 /0.9972 | 129470 /0.4939 | 51.9280 /0.9973 | 126006 /0.4807 | 52.4411 /0.9975 | 92044 /0.3511 | 55.7068 /0.9988 |
| Baboon | 131048 /0.4999 | 51.1409 /0.9991 | 131004 /0.4997 | 51.3619 /0.9991 | 130749 /0.4988 | 51.6104 /0.9992 | 130248 /0.4969 | 51.8469 /0.9992 | 129530 /0.4941 | 52.0385 /0.9992 | 126808 /0.4837 | 52.4461 /0.9992 | 93257 /0.3557 | 55.6309 /0.9996 |
| Crowd | 129529 /0.4941 | 50.1606 /0.9980 | 129386 /0.4936 | 50.9942 /0.9981 | 128951 /0.4919 | 51.4358 /0.9982 | 127977 /0.4882 | 51.8742 /0.9982 | 126471 /0.4824 | 52.2819 /0.9983 | 121834 /0.4648 | 53.0444 /0.9984 | 98264 /0.3748 | 55.0397 /0.9989 |
| Plane | 131069 /0.5000 | 45.3539 /0.9939 | 130804 /0.49898 | 51.7207 /0.99768 | 130541 /0.49797 | 52.0452 /0.9977 | 130004 /0.4959 | 52.3245 /0.9977 | 129085 /0.4924 | 52.5890 /0.9978 | 126181 /0.4817 | 53.0896 /0.9979 | 105403 /0.4021 | 55.0693 /0.9984 |
| Peppers | 131064 /0.5000 | 50.4753 /0.9971 | 130980 /0.4996 | 50.9755 /0.9972 | 130591 /0.4982 | 51.3869 /0.9973 | 129939 /0.4957 | 51.6763 /0.9974 | 128872 /0.4916 | 51.9267 /0.99754 | 125136 /0.4774 | 52.4908 /0.9977 | 91590/ 0.3494 | 55.7243 /0.9989 |
| Tank | 131072 /0.5000 | 51.5168 /0.9981 | 130999 /0.4997 | 51.8410 /0.9982 | 130780 /0.4989 | 52.1083 /0.9983 | 130434 /0.4976 | 52.3087 /0.9983 | 129047 /0.4923 | 52.7114 /0.9985 | 122289 /0.4665 | 54.0191 /0.9988 | 115805 /0.4418 | 54.6884 /0.9990 |

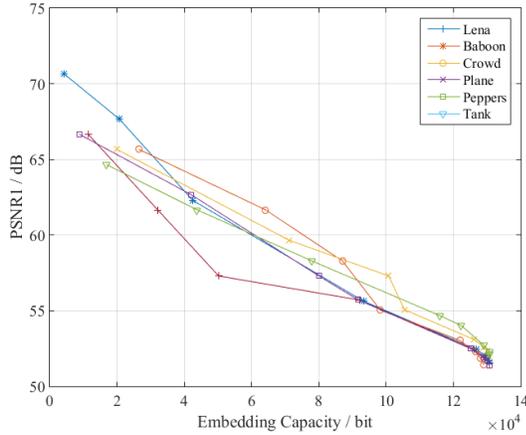

Fig. 7. Relationships of $PSNR_1$ (dB) versus EC (bit) on six different images.

*1) DE-SBED*

*a) Reversibility of plaintext recovery*

Data embedding is implemented in public servers while decryption and plaintext recovery can only be performed by the secret key owner. The directly decrypted result of the marked ciphertext is the marked plaintext whose PSNR was calculated and recorded as $PSNR_1$. The EC, ER, $PSNR_1$, SSIM with different $h_{fid}$ on six test images are recorded in Table IV. The maximum EC of an image is determined by the number of the available pixel pairs which meet the constraints in Eqs. (10)-(12). Due to the introduction of PVO, the values of adjacent pixels tend to be equal. Therefore, the number of pixel pairs that satisfy the constraints can be effectively increased, thereby improving the embedding capacity. On the other hand, the larger $h_{fid}$ is set, the more pixel pairs satisfy the constraints, and the higher the embedding rate (ER) is. When $h_{fid}=\infty$, the ER of an image reaches the maximum. Table IV shows that when $h_{fid} \geqslant 5$, the ER could reach or approach 0.5bpp that is the theoretical maximum value of DE algorithm.

To further test the distortion in the direct decrypted result, we tested the distortion of 1000 selected images from USC-SIPI. The experimental results show that the marked images and the original images cannot be visually distinguished. When $1 \leqslant h_{fid} \leqslant 10$, all the maximum ER of the 1000 images remains above 0.45bpp, and the corresponding $PSNR_1$ reaches an average of 52.6773dB. When $h_{fid} = 0$, the maximum ER of the 1000 images remains above 0.35bpp, and the corresponding $PSNR_1$ reaches 55.1270dB on average. Fig. 7 presents the relationships of $PSNR_1$ versus EC in directly decrypted images for six different test images.

We continue to make a comparison of $PSNR_1$ among the proposed PKRDH-ED1, existing representative RDH-ED algorithms [3][21][29][31][32] and Tian' algorithm [35] under different embedding rates. The results of different images from USC-SIPI show that the proposed algorithm have a better fidelity of the directly decrypted results from the marked ciphertext. In Fig. 8, we demonstrated the comparison results of $PSNR_1$ from Lena (Fig. 8(a)) and Plane (Fig. 8(b)).

*DE recovery* is then performed on the marked plaintext to obtain the recovered images whose PSNR was calculated and recorded as $PSNR_2$. *DE recovery in single bit encrypted domain* was performed by public servers to obtain a newly restored ciphertext which can be decrypted by the user. The PSNRs of the decrypted results of the newly restored ciphertext were calculated and recorded as $PSNR_3$. In our results, all the values of $PSNR_2$ and $PSNR_3$ are "∞", which demonstrates that no distortion is reminded in recovered images.

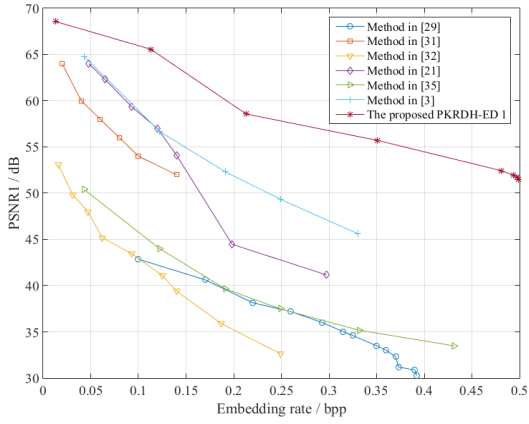
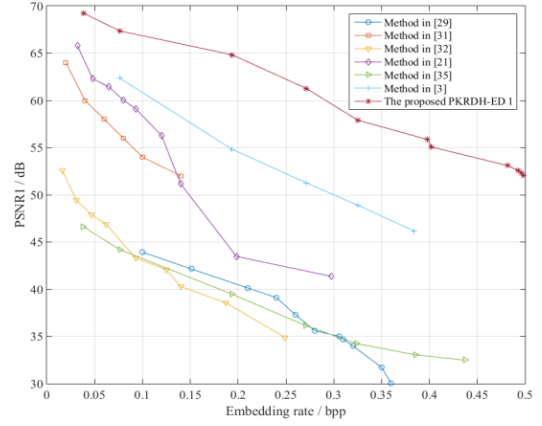

Fig. 8. The comparison of PSNR1 (dB) with different ER (bpp) on (a) Lena; (b) Plane.

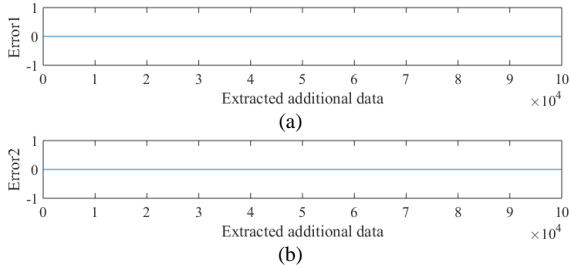

Fig. 9. Errors of the extracted data: (a) Error1: errors of extraction in case *a*; (b) Error2: errors of extraction in case *b*.

TABLE V
THE SETTING OF $\alpha$ AND $N$ FOR REVERSIBILITY OF LWE DECRYPTION

| $N$ | 240 | 280 | 380 | 820 |
|---|---|---|---|---|
| $\alpha_{min}/\times 10^{-4}$ | 5.3791 | 4.2686 | 2.6998 | 8.5169 |
| $N_{max}$ | 1 | $\text{Log}_2 3$ | 2 | 3 |
| $\alpha_{max}/\times 10^{-4}$ | 7.4714 | 6.5671 | 5.1132 | 33.0304 |
| $PSNR_1$ | $\infty$ | $\infty$ | $\infty$ | $\infty$ |

*a) Accuracy of data extraction*

Data extraction could be operated only by the secret key owner. There are two cases of data extraction: *a)* the user extracts data from the marked plaintext by using DE extraction; *b)* the user decrypts the encrypted additional data to obtain the embedded data via the secret key. We have calculated the differences of the extracted bits and the to-be-embedded additional bits for $10^4$ times in the mentioned two cases. As shown in Fig. 9, the results of extraction accuracy in the two cases are both 100%.

*2) PKR-ER*
*a) Reversibility of plaintext recovery*

In PKR-ER, $|e^T \cdot a_r|$ is constrained by the setting of $\alpha$. Then we could embed an $N$-bit additional data $m_e \in \mathbb{Z}_{2^N}$ into the ciphertext encrypted from one bit of plaintext by changing the location of $\lambda$ within its redundant region. The ER of the proposed algorithm 2 is $N$ bits per bit (bpb) of plaintext. According to Eq. (26), the interval of $\lambda$ after embedding is $[0, (2^N-1)Q_{step}]$. To ensure no overflow is resulted on the region of $\lambda$, the following condition should be satisfied [19][20]:

$$|e^T \cdot a_r| + (2^N-1) Q_{step} < q/4 \quad (29)$$

Namely, the necessary and sufficient condition of correct decryption of PKRDH-ED algorithm 2 is obtained:

$$|e^T \cdot a_r| < \lfloor q/2^{N+2} \rfloor \quad (30)$$

In [19][20], it has been deduced that $e^T \cdot a_r$ follows the Gaussian distribution of $N(0, \sqrt{d/2} \cdot \alpha)$. And the probability of distribution function of $|e^T \cdot a_r|$ was also obtain. Then we can have the decryption-error probability of PKR-ER:

$$P(|e^T a_r| \geq \frac{q}{2^{N+2}}) \leq \frac{2^{N+2} \alpha}{q} \cdot \sqrt{\frac{d}{\pi}} \exp(-\frac{q^2}{2^N d \alpha^2}) \quad (31)$$

According to Eq. (31), when $\alpha$ was fixed, the smaller $N$ is set, the less probably decryption errors might occur. Meanwhile, the ER decreases. Once $N$ was fixed, the smaller $\alpha$ is set, the better the reversibility would be ensured. Then, ER is $N$ bpb. However, if $\alpha$ is too small, the security of LWE encryption might be seriously compromised [20]. According to [20][36], schemes based on LWE problem generally require $\alpha q > 2\sqrt{n}$, which provide us the minimum value of $\alpha$ for PKR-ER, *i.e.*, $\alpha_{min} = 2\sqrt{n}/q$.

We can obtain the maximum value of $N$ on the condition that $\alpha$ is set $\alpha_{min}$ by experiments, recorded as $N_{max}$ in Table V. Then the maximum value of $\alpha$ recorded as $\alpha_{max}$ could also be obtained by setting $N=1$ in experiments. In the experiments, $10^6$ bits of decryption were implemented and no error occurred, which verified the availability of the parameters $\alpha_{max}$, $N_{max}$.

In the experiments, PEK is necessary for embedding and accessible in untrusted public environments. Considering the efficiency and security requirements, the parameters are set: $n=240$, $\alpha \in (\alpha_{min}, \alpha_{max})$, $N=1$. Only the secret key owner can decrypt the marked ciphertext, and the values of $PSNR_1$ of test images are all "$\infty$". Different from DE-SBED, there is no distortion in the results directly decrypted from the marked ciphertext and the recovery operation is not required.

*b) Accuracy of data extraction*

Data extraction can only be implemented by the user in private. According to the methodology in Section III. C, it is $|e^T \cdot a_r| < Q_{step}$ that is the necessary and sufficient condition to ensure the accuracy of data extraction.

In the experiments, we have also recorded the differences of the extracted bits and the to-be-embedded bits by setting $\alpha$

between ($\alpha_{min}$, $\alpha_{max}$). The results demonstrated that the accuracy is 100%.

*B. Security*

Security of PKRDH-ED includes three aspects: *a)* the one-way attribute in public embedding key (PEK) generation. PEK should not reveal any information of deducing the secret key or extraction key. *b)* Consistent security of embedding. The embedding operations should not weaken the security of the original encryption. *c)* The confidentiality of the embedded information. The embedded information cannot be obtained by an attacker without the extraction key.

*1) DE-SBED*

*The one-way attribute*: The public information about embedding includes the public key and parameters settings while the secret information about extraction is only the secret key. According to the security principle of public key cryptosystem [38]-[40], the secret key cannot be obtained by the public information, thus ensuring the one-way attribute of public information.

*Consistent security of embedding*: The operations in DE-SBED mainly consist of bits encryption of LWE, ciphertext position shift or replacement. All the operations are based on normal encryption and would not reveal anything about the secret key or reduce the security of LWE encryption [3], thus realizing the consistent security of embedding.

*The confidentiality of the embedded information*: The additional data is encrypted by LWE encryption before embedding by a data hider in public, which could ensure the secrecy of the additional data during the transmission. Due to the random variable $a_r$ in Eqs. (18-19), different ciphertext would be independent from each other even if they are encrypted from the same plaintext by the same public key. Therefore, the confidentiality of the embedded information is ensured.

Above all, a third party or an attacker in public environments cannot gain any information about the secret key or plaintext, thus ensuring the security of PKRDH-ED.

*2) PKR-ER*

*The one-way attribute*: Since PEK is constructed here, we deduce the probability distribution function (PDF) of PEK to demonstrate the one-way attribute of PEK generation in this section.

$Q_{step}$ is calculated by $N$ and $q$, which are uncorrelated with the secret key or plaintext. The probability of $\gamma$ is denoted as $P_\gamma$. The probability of $\lambda$ is denoted as $P_\lambda$. According to Eq. (24), sign factor $\gamma$ is determined by the distribution of $\lambda$:

$$P_\gamma(\gamma=+1) = P_\lambda(\lambda \in [0, \lfloor q/4 \rfloor) \cup [\lfloor q/2 \rfloor, \lfloor 3q/4 \rfloor)) \quad (32)$$

$$P_\gamma(\gamma=-1) = P_\lambda(\lambda \in [\lfloor q/4 \rfloor, \lfloor q/2 \rfloor) \cup [\lfloor 3q/4 \rfloor, q)) \quad (33)$$

In Eq. (23), $\lambda = e^T \cdot a_r + m \cdot \lfloor q/2 \rfloor$. $e^T \cdot a_r \in [0, \lfloor q/4 \rfloor) \cup [\lfloor 3q/4 \rfloor, q)$ according to Eq. (30).

Then we can deduce the Eqs. (34-35) from Eqs. (32-33):

$$P_\gamma(\gamma=+1) = P_\lambda(e^T \cdot a_r \in [0, \lfloor q/4 \rfloor)) \quad (34)$$

$$P_\gamma(\gamma=-1) = P_\lambda(e^T \cdot a_r \in [\lfloor 3q/4 \rfloor, q)) \quad (35)$$

Since $e^T \cdot a_r$ follows a Gaussian distribution with a mean of 0, $P_\lambda(e^T \cdot a_r \in [0, \lfloor q/4 \rfloor)) = P_\lambda(e^T \cdot a_r \in [\lfloor 3q/4 \rfloor, q)) = 1/2$ based on the symmetry of the Gaussian distribution.

Then we can obtain the probability $P_\gamma$:

$$P_\gamma(\gamma=+1) = P_\gamma(\gamma=-1) = 1/2 \quad (36)$$

To test the correctness of Eq. (36), we have encrypted more than $10^6$ bits of data to obtain the distribution of $e^T \cdot a_r$ and embedded randomly sampled data to obtain $\lambda$ before and after embedding. Fig. 10 shows the symmetry of the distributions of $e^T \cdot a_r$ and $\lambda$ before and after embedding.

Therefore, $\gamma$ is proved to be randomly distributed and would not reveal anything about the secret key or plaintext.

*Consistent security of embedding*: Assuming that the ciphertext follows the uniform distribution, *i.e.*, $c \sim U(0,q)$ [19][20], it can be deduced that the PDF of the marked ciphertext is also the uniform distribution *i.e.*, $c' \sim U(0,q)$ according to Eqs. (26, 36): $c' = c + m_e \times \gamma Q_{step}$ and $P_\gamma(\gamma=+1) = P_\gamma(\gamma=-1) = 1/2$. It was proved that the distribution of ciphertext before and after embedding has not changed.

Statistic features of histogram, information entropies of ciphertext before and after embedding were obtained by experiments. We tested four groups of sample data to obtain histograms and the average information entropies of the ciphertext before and after embedding. Fig. 11 demonstrates the results of histograms for $n=240$, $q=57601$, $\alpha \in (\alpha_{min}, \alpha_{max})$, $N=1$. There were $1.92 \times 10^7$ bits of data sampled as plaintext in each group.

The average information entropies of the original ciphertext and the marked ciphertext were denoted as $H$ and $H'$ in Table VI. The theoretical ideal maximum entropy in Galois field $\mathbb{Z}_q$ is denoted as $H_{ideal}$, $H_{ideal} = -q \times (1/q) \times \log_2(q^{-1}) = 15.8138$ when $q=57601$.

The experimental results demonstrate that the histograms have not changed significantly after embedding. The recoding of embedding on the ciphertext is equivalent to coarse random scrambling, which could contribute to the encryption, so the average information entropy of the marked ciphertext is not less than the original one. Therefore, theoretical analysis and experimental results both demonstrate that the consistent security of embedding can be ensured.

*The confidentiality of the embedded information*: Data extraction can be implemented only by obtaining the quantization variable $\lambda$. $\lambda$ can only be calculated by using the secret key, which is determined by the principle of LWE encryption [38]-[40]. Therefore, the confidentiality of the embedded information can be ensured.

*C. Efficiency*

*1) Computational complexity*

As discussed in Section III, DE-SBED is optimized based on the algorithm in [3]. Though the application framework of DE-SBED is the same as FHEE-ED in [3], the methodology of the realization is different. No homomorphic addition/ multiplication or bootstrapping are introduced while the performance of homomorphic operations is achieved. The operations of embedding in DE-SBED just consist of ciphertext position shift or replacement based on bit operations of DE, thus resulting in a high operation speed of embedding. Key consumption is also reduced from $O(100)$-$O(10)$ in [3] to $O(1)$.

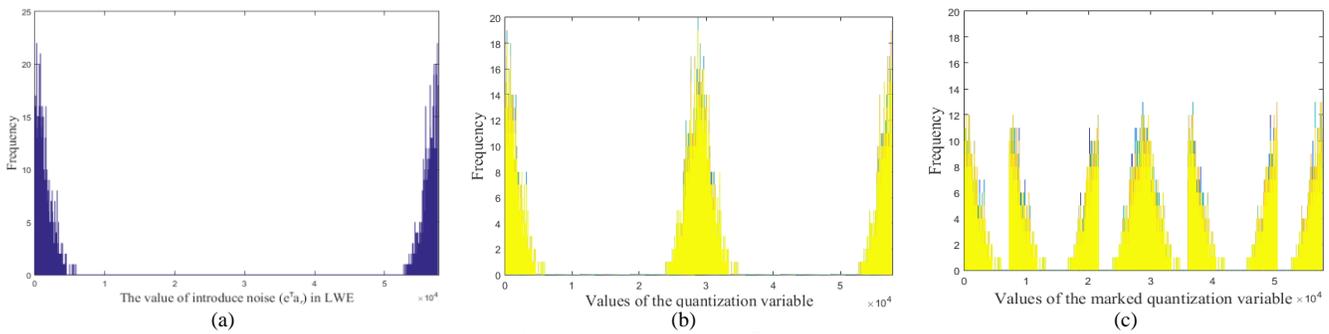

Fig. 10. The distributions of different variables with $1.92\times10^6$ bits of sampled data: (a) $e^T\cdot a_r$; (b) $\lambda$ before embedding; (c) marked $\lambda$ after embedding.

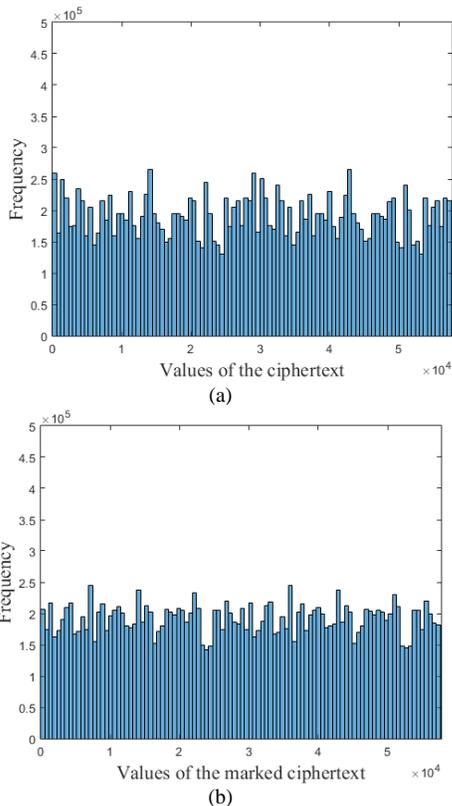

Fig. 11. The distributions of ciphertext with $1.92\times10^7$ bits of sampled data: (a) before embedding; (b) after embedding.

TABLE VI
THE AVERAGE INFORMATION ENTROPIES OF CIPHERTEXT

| Group No. | 1 | 2 | 3 | 4 |
|---|---|---|---|---|
| $H$ | 12.6462 | 12.7593 | 11.8302 | 11.8366 |
| $H'$ | 12.7922 | 12.8191 | 12.7640 | 12.7777 |

TABLE VII
COMPARISON OF COMPLEXITY

| Main operation | Typical schemes | Complexity |
|---|---|---|
| Stream encryption | [4][6][22][23][25] | $O(x)$ |
| Stream encryption with preprocessing | [1][5][7][24] | $O(x\log_2 x)$-$O(x^2)$ |
| Paillier encryption | [26]-[32] | $O(x^3)$ |
| **LWE encryption** | [19][20], **the proposed** | $\mathbf{O(x^2)}$ |
| FHE encryption | [3] | $O(x^2)$- $O(x^3)$ |
| Bootstrapping | [3] | $O(x^3)$ |
| Secret sharing | [18] | $O(x)$ |

TABLE VIII
ELAPSED TIME ($ms$) OF OPERATIONS ON $2.4\times10^3$ BYTES

| Operation | Encryption | Decryption | Embedding | Extraction |
|---|---|---|---|---|
| Algorithm 1 | 24.732 | 27.728 | 1.344 | 1.027 |
| Algorithm 2 | 24.732 | 27.728 | 1.053 | 9.160 |

Compared with Paillier encryption or elliptic curve encryption, LWE encryption has a higher computational speed due to its brief structure and linear operations. Let $x$ be the length of plaintext, the computational complexities of embedding and extraction in DE-SBED are $O(1)$ while the complexity of PVO processes is $O(x\log_2 x)$-$O(x^2)$. As for PKR-ER, the computational complexities of embedding is $O(1)$ while the extraction is $O(x^2)$.

The computational complexities of different RDH-ED schemes are compared in Table VII, which demonstrates that the proposed two algorithms have a low computation complexity compared with other public key encryption based RDH-ED algorithms. Table VIII lists the elapsed time of the operations in the proposed algorithms. The elapsed time is the time (milliseconds, $ms$) for encrypting/ decrypting $2.4\times10^3$ bytes from a $512\times512$ image, or embedding/ extraction at the rate of 0.5bpp. The results show that the proposed two algorithms are practicable in application.

*2) Embedding Rates*

In DE-SBED, the introduction of POV contributes to the fidelity and EC of RDH-ED. According to the results in Section IV. A. 1, the ER of different test images can reach or approach the theoretical maximum ER of DE, 0.5bpp. Meanwhile, under the same ER, the distortions in the directly decrypted images are smaller than other existing RDH-ED methods.

PKR-ER is implemented based on the redundancy in encryption. It has nothing to do with the type of medium or the content of the plaintext. According to the results in Section IV. A. 2, the ER is $N$ bpb ($8N$ bpp when the plaintext bits are from 8-bit image pixels). $N$ effects the setting of the quantization step and the constraints on $\alpha$. The maximum of $N$ is determined by the parameters $n$ and $\alpha_{min}$. According to the results in Table V, the embedding rates has achieved 1bpb or higher by setting $n\geq 240$, $\alpha\in(\alpha_{min}, \alpha_{max})$.

*3) Storage cost*

There is no extra storage cost brought in by the operations of embedding. The storage cost is mainly resulted from the cryptosystems. The public key encryption algorithms, including the Paillier algorithm and the LWE algorithm, have ciphertext extension. It is resulted by the principle of mathematics of public key encryption and could provide reliable security guarantees for secret communication. In [20], the ciphertext extension of Paillier and LWE encryption was discussed in detail. The extension of LWE could reach $O(n^2 \log n)$ while Paillier is about $O(n^3)$.

Since ciphertext is usually stored in the server or the cloud in the application of PKRDH-ED, the local storage cost of users is not too much. In practice, we concentrate more on the elapsed time or computational complexity of encryption, decryption, embedding, and data extraction, which is more important to the efficiency and has been discussed in Section IV. C. 1. On the other hand, ciphertext extension has provided considerable redundancy in encryption which could be taken advantage of for embedding.

## V. Conclusion

Considering the application prospects of RDH-ED in distributed deep learning privacy and secure MPC systems, the public key embedding mechanism of RDH-ED is proposed in this paper. We have discussed the characteristics and security requirements of PKRDH-ED by analyzing its application in federated learning system.

Two algorithms with PEK mechanism are proposed based on LWE: In DE-SBED, we construct PKE mechanism based on the characteristics of single bit encryption of LWE. It supports any untrusted third party in public to embed additional data with a high running speed. By introducing PVO before encryption, the direct decryption distortion of the marked ciphertext is effectively reduced. PKR-ER is based on the redundancy in encryption resulted by the probabilistic decryption of LWE. Through the quantization on the encrypted domain and recoding on the ciphertext, public embedding key is constructed which is independent from the public key for encryption. The decryption and extraction processes are implemented based on different quantization rules. Therefore, there is no distortion in the direct decrypted result of the marked ciphertext. ER of PKR-ER can reach more than 1bpb. Theoretical analysis and experimental results demonstrate the performance in correctness, security and efficiency of the proposed algorithms. Specifically, the one-way attribute of public information is of importance in the security of PKRDH-ED.

Future investigation will focus on optimizing the technique of PKRDH-ED to further improve the efficiency, and applying this technique to distributed computing applications, such as federated learning, secure MPC.

## VI. Acknowledgements

This work was supported by the National Natural Science Foundation of China under Grant No. 62102450, Grant No. 61872384.

## References

[1] K. Ma, W. Zhang, X. Zhao, N. Yu, and F. Li, "Reversible data hiding in encrypted images by reserving room before encryption," *IEEE Trans. Inf. Forensics Security*, vol. 8, no. 3, pp. 553–562, Mar. 2013.

[2] Y.-Q. Shi, X. Li, X. Zhang, H. Wu, "Reversible Data Hiding: Advances in the Past Two Decades." *IEEE Access*, vol.4, no. 5, pp. 3210-3237, May. 2016.

[3] Y. Ke, M. Q. Zhang, J. Liu, T. T. Su, X. Y. Yang, "Fully Homomorphic Encryption Encapsulated Difference Expansion for Reversible Data hiding in Encrypted Domain," *IEEE Transactions on Circuits and Systems for Video Technology*, vol. 30, no. 8, pp. 2353–2365, Aug. 2020.

[4] X. Zhang, "Reversible data hiding in encrypted image," in *IEEE Signal Processing Letters*, vol.18, no. 4, pp. 255-258, Apr. 2011.

[5] J. Zhou, W. Sun, L. Dong, X. Liu, O. C. Au, and Y. Y. Tang, "Secure reversible image data hiding over encrypted domain via key modulation," *IEEE Trans. Circuits Syst. Video Technol.*, vol. 26, no. 3, pp. 441–452, Mar. 2016.

[6] X. Wu and W. Sun, "High-capacity reversible data hiding in encrypted images by prediction error," *Signal Processing*, vol.104, no. 11, pp. 387-400, Nov. 2014.

[7] Z. Qian, X. Zhang, S Wang, "Reversible data hiding in encrypted JPEG bitstream," *IEEE Transaction on Multimedia*, vol.16, no. 5, pp. 1486-1491, May. 2014.

[8] Z. Yin, Y. Peng and Y. Xiang, "Reversible Data Hiding in Encrypted Images Based on Pixel Prediction and Bit-plane Compression," *IEEE Transactions on Dependable and Secure Computing*, doi: 10.1109/TDSC.2020.3019490.

[9] W. Puech, M. Chaumont, and O. Strauss, "A reversible data hiding method for encrypted images," in Proc. SPIE 6819, Security, Forensics, Steganography, and Watermarking of Multimedia Contents X, 2008, pp. 68 191E–68 191E–9.

[10] W. Zhang, K. Ma, and N. Yu, "Reversibility improved data hiding in encrypted images," *Signal Processing*, vol. 94, no. 1, pp. 118–127, 2014.

[11] M. Li, D. Xiao, Y. Zhang, H. Nan, "Reversible data hiding in encrypted images using cross division and additive homomorphism", *Signal Processing: Image Communication*, vol. 39PA, no. 11, pp. 234–248, 2015.

[12] Jakub Konecný, H. Brendan McMahan, Felix X. Yu, Peter Richtárik, Ananda Theertha Suresh, and Dave Bacon. 2016. Federated Learning: Strategies for Improving Communication Efficiency. *CoRR abs/1610.05492 (2016)*. arXiv:1610.05492 http://arxiv.org/abs/1610.05492.

[13] Yang Q., Liu Y., Chen T., et al. "Federated Machine Learning: Concept and Applications". *ACM Transactions on Intelligent Systems and Technology*, vol.10, no. 2, pp. 1-19, Feb. 2019.

[14] Le Trieu Phong, Yoshinori Aono, Takuya Hayashi, Lihua Wang, and Shiho Moriai. 2018. Privacy-Preserving Deep Learning via Additively Homomorphic Encryption. *IEEE Trans. Information Forensics and Security.* 13, 5 (2018), 1333–1345.

[15] Cao, D., et al. "Understanding Distributed Poisoning Attack in Federated Learning." *2019 IEEE 25th International Conference on Parallel and Distributed Systems (ICPADS) IEEE*, 2019.

[16] N. Papernot, P. McDaniel, S. Jha, M. Fredrikson, Z. B. Celik and A. Swami, "The Limitations of Deep Learning in Adversarial Settings," *2016 IEEE European Symposium on*

Security and Privacy (EuroS&P), 2016, pp. 372-387, doi: 10.1109/EuroSP.2016.36.

[17] Chulin Xie, Keli Huang, Pin-Yu Chen, Bo Li. "DBA: Distributed Backdoor Attacks against Federated Learning," *Proc of the 7th Int Conf on Learning Representations*. 2019 [2021].

[18] Y. Ke, M. Zhang, X. Zhang, et. al., "A Reversible Data hiding Scheme in Encrypted Domain for Secret Image Sharing based on Chinese Remainder Theorem," *IEEE Trans. Circuits and Systems for Video Technology*, doi: 10.1109/TCSVT. 2021. 3081575.

[19] Y. Ke, M. Zhang, J. Liu, "Separable multiple bits reversible data hiding in encrypted domain," in *Digital Forensics and Watermarking - 15th International Workshop, IWDW 2016*, Beijing, China, LNCS, 10082, pp. 470-484, 2016.

[20] Y. Ke, M. Zhang, J. Liu, T. Su, et. al. "A multilevel reversible data hiding scheme in encrypted domain based on LWE," *Journal of Visual Communication & Image Representation*, vol. 54, no. 7, pp. 133-144, 2018.

[21] Li Z X, Dong D P, Xia Z H, et al, "High-capacity reversible data hiding for encrypted multimedia data with somewhat homomorphic encryption", *IEEE Access*, vol.6, no.10, pp. 60635-60644, 2018.

[22] X. Zhang, "Separable reversible data hiding in encrypted image," *IEEE Transactions on Information Forensics and Security*, vol.7, no. 2, pp. 826-832, 2012.

[23] H.-Z. Wu, Y.-Q. Shi, H.-X. Wang, et al, "Separable reversible data hiding for encrypted palette images with color partitioning and flipping verification", *IEEE Transactions on Circuits and Systems for Video Technology*, vol.27, no. 8, pp. 1620 - 1631, 2016.

[24] P. Puteaux and W. Puech, "An efficient msb prediction-based method for high-capacity reversible data hiding in encrypted images," *IEEE Trans. Inf. Forensics Secur.*, vol. 13, no. 7, pp. 1670 - 1681, 2018.

[25] F. J. Huang, J. W. Huang and Y. Q. Shi, "New Framework for Reversible Data Hiding in Encrypted Domain," *IEEE Transactions on information forensics and security*, vol. 11, no. 12, pp. 2777-2789, Dec. 2016.

[26] Y. -C. CHEN, C. -W. SHIU, G. HORNG. "Encrypted signal-based reversible data hiding with public key cryptosystem," *Journal of Visual Communication and Image Representation*, vol. 25, no. 5, pp.1164-1170, 2014.

[27] C. -W. Shiu, Y. -C. Chen, and W. Hong, "Encrypted image-based reversible data hiding with public key cryptography from difference expansion," *Signal Processing: Image Communication*, vol. 39, pp. 226–233, 2015.

[28] X. Wu, B. Chen, and J. Weng, "Reversible data hiding for encrypted signals by homomorphic encryption and signal energy transfer," *Journal of Visual Communication and Image Representation*, vol. 41, no. 11, pp. 58–64, 2016.

[29] X. -P. Zhang, J Loong, Z Wang, et al, "Lossless and reversible data hiding in encrypted images with public key cryptography", *IEEE Transactions on Circuits and Systems for Video Technology*, vol. 26, no. 9, pp. 1622 – 1631, 2016.

[30] H.-T. Wu, Y.-M. Cheung, J. -W. Huang, "Reversible data hiding in paillier cryptosystem", *Journal of Visual Communication and Image Representation,* vol. 40, no. 10, pp. 765-771, 2016.

[31] LI M, LI Y., "Histogram shifting in encrypted images with public key cryptosystem for reversible data hiding", *Signal Process*, vol. 130, no. 1,pp: 190-196, 2017.

[32] S. -J. Xiang, X. Luo, "Reversible Data Hiding in Homomorphic Encrypted Domain By Mirroring Ciphertext Group", *IEEE Trans. Circuits Syst. Video Technol.* vol. 28, no. 11,pp: 3099-3110, 2018.

[33] Puteaux P, Ong S Y, Wong K S, et al. "A Survey of Reversible Data Hiding in Encrypted Images-The First 12 Years". *Journal of Visual Communication and Image Representation*, vol. 77, no. 5, 2021: 103085.

[34] Z. Ni, Y.-Q. Shi, N. Ansari and W. Su, "Reversible data hiding," *IEEE Trans. Circuits Syst. Video Technol.*, vol. 16, no. 3, pp. 354–362, Mar. 2006.

[35] J. Tian, "Reversible data embedding using a difference expansion," *IEEE Trans. Circuits Syst. Video Technol.*, vol. 13, no. 8, pp. 890–896, Aug. 2003.

[36] O. Regev. "On lattices, learning with errors, random linear codes and cryptography," *Journal of the ACM*, vol.56, no. 6, pp.34, Jun. 2009.

[37] O. Regev. "The learning with errors problem," in *Proc of Int Conf on Public Key Cryptography* (PKC2007), Berlin, Germany, 2007, pp. 315-329.

[38] D. Micciancio, O. Regev, "Lattice-based Cryptography", in *Post-quantum Cryptography*, D. J. Bernstein and J. Buchmann (eds.), Berlin, Heidelberg, Germany: Springer, 2008. pp. 147-191.

[39] Nicolas Gama, Phong Q. Nguyen, "Predicting lattice reduction", in *Advances in cryptology-Eurocrypt2010: 27th Annual International Conference on the Theory and Applications of Cryptographic Techniques.* Istanbul, Turkey, pp. 31-51, 2008.

[40] M Ruckert, M Schneider. "Estimating the security of latticed-based cryptosystems", (2010) [Online] Available: http:// eprint.icur.org/ 2010/ 137.pdf.